\documentclass[manuscript]{aastex}

\usepackage{graphicx}
\usepackage{amssymb}
\usepackage{natbib}
\usepackage{amsmath}
\usepackage{url}

\slugcomment{}

\shorttitle{First observations of a dome-shaped large-scale coronal EUV wave}
\shortauthors{Veronig et al.}

\begin{document}

\title{First observations of a dome-shaped large-scale coronal EUV wave}

\author{A.M. Veronig}
\affil{Institute of Physics, University of Graz,
    Universit\"atsplatz 5, A-8010 Graz, Austria}
    \email{asv@igam.uni-graz.at}

\author{N. Muhr}
\affil{Institute of Physics, University of Graz,
    Universit\"atsplatz 5, A-8010 Graz, Austria}

\author{I.W. Kienreich}
\affil{Institute of Physics, University of Graz,
    Universit\"atsplatz 5, A-8010 Graz, Austria}

\author{M. Temmer}
\affil{Institute of Physics, University of Graz,
    Universit\"atsplatz 5, A-8010 Graz, Austria;
    Space Research Institute, Austrian Academy of Sciences, Schmiedlstra{\ss}e 6, A-8042 Graz, Austria
    }

\author{B. Vr\v{s}nak}
\affil{Hvar Observatory, Faculty of Geodesy, Ka\v{c}i\'ceva 26, 1000 Zagreb, Croatia}

\begin{abstract}
We present first observations of
a dome-shaped large-scale EUV coronal wave, recorded by
the EUVI instrument onboard STEREO-B on January 17, 2010. The main arguments that the
observed structure is the wave dome (and not the CME) are: a) the spherical form and sharpness of
the dome's outer edge and the erupting CME loops observed inside the dome;  b) the low-coronal wave
signatures above the limb perfectly connecting to the on-disk signatures of the wave; c) the
lateral extent of the expanding dome which is much larger than that of the coronal dimming; d) the
associated high-frequency type II burst indicating shock formation low in the corona. The velocity
of the upward expansion of the wave dome ($v \sim 650$~km~s$^{-1}$) is larger than that of the
lateral expansion of the wave ($v \sim 280$~km~s$^{-1}$), indicating that the upward dome expansion
is driven all the time, and thus depends on the CME speed, whereas in the lateral direction it is
freely propagating after the CME lateral expansion stops.
We also examine the evolution of the perturbation characteristics: 
First the perturbation profile steepens and the amplitude increases. Thereafter, the amplitude
decreases with r$^{-2.5 \pm 0.3}$, the width broadens, and the integral below the perturbation 
remains constant. Our findings are consistent with the spherical expansion and decay of a weakly shocked 
fast-mode MHD wave.

\end{abstract}


\keywords{Sun: corona --- Sun: coronal mass ejections (CMEs) --- Sun: flares}

\section{Introduction}

Large-scale propagating disturbances in the solar atmosphere occuring in association with
flares and coronal mass ejections (CMEs) have been first observed in chromospheric filtergrams \citep{moreton60,athay61}. These
``Moreton waves'' propagate with typical velocities in the order of 1000~km~s$^{-1}$, and have been
interpreted as the ground-track of a dome-shaped MHD wave front propagating through the solar corona, which
compresses and pushes the chromospheric plasma downward when sweeping over it \citep{uchida68}. 
The Extreme-ultraviolet Imaging Telescope \citep[EIT;][]{delaboudiniere95} onboard SOHO for the first time
imaged such wave-like disturbances in coronal emission lines \citep{moses97,thompson98}.

EIT waves have been initially interpreted as the coronal
counterparts of the chromospheric Moreton waves as predicted in Uchida's fast-mode coronal MHD wave model \citep{thompson99}.  
However, differences in morphology and propagation velocities of Moreton waves and EIT waves, which lie mostly in the range 200--400 km~s$^{-1}$
\citep{klassen00,thompson09} led to severe doubts of this interpretation and alternative models were put forward.
Some of them question if the phenomenon is a wave at all, and instead suggest that it is a signature of the large-scale coronal restructuring
due to the erupting CME causing plasma compression or localized energy release
\citep[e.g.][]{delanee99,chen02,attrill07}. For detailed discussions of the different models, 
we refer to the recent reviews by \cite{warmuth07}, \cite{vrsnak08}, and \cite{wills10}.

There seems to be some consensus that Moreton waves are indeed shock waves, as is suggested by their high propagation velocities and perturbation characteristics.
It was shown that the amplitude and velocity of Moreton waves decrease, and the width of the wave pulse broadens as it propagates,
consistent with a large-amplitude wave or freely propagating shock wave that formed by steepening of a simple wave \citep[e.g.][]{warmuth04b}.
However, the situation is unclear for large-scale waves observed in the corona. There is rather qualitative
insight that the wave fronts not only become more diffuse but also broaden as they propagate \citep{thompson99,klassen00,podladchikova05},
which may be due to energy flux conservation or dispersion of the wave  \citep[e.g.][]{wills10}. On a statistical basis of different EIT waves, \cite{warmuth10}
concludes that for larger distances, the perturbation amplitudes tend to become smaller and the width larger.
However, a case study of the wave evolution in high-cadence TRACE images by \cite{wills06} revealed that the 
width of the wave pulse remained constant during the propagation.

Due to the low EIT cadence of $\sim$12--15~min it was not possible to detect fast EIT waves and
to study the evolution of the wave pulse characteristics, which is important in constraining
the physical processes. The Extreme Ultraviolet Imager \citep[EUVI;][]{howard08} instruments onboard the
twin spacecraft of the Solar-Terrestrial Relations Observatory
\citep[STEREO;][]{kaiser08} overcome these limitations and offer several advantages for the study of large-scale coronal waves, in particular
due to their high cadence, large field-of-view (FoV), high sensitivity, and the simultaneous observations from two vantage points.

Since the launch of the STEREO satellites in October 2006, a variety of case studies of large-scale coronal waves have been carried out using EUVI data \citep{long08,veronig08,gopalswamy09,patsourakos09,patsourakos09b,attrill09,cohen09,kienreich09,ma09,zhukov09,dai10}.
The typical propagation velocities that were derived from EUVI waves (that all occurred during solar minimum conditions and were not accompanied by Moreton waves) 
are in the range 200--350~km~s$^{-1}$. 
For the kinematical evolution, the results so far are rather inconclusive: some of the waves studied with EUVI showed evidence for
deceleration during their propagation, consistent with the decay and deceleration of a large amplitude wave to
the fast-mode speed of the ambient corona \citep[e.g.][]{long08,veronig08}, whereas
others showed constant velocity \citep[e.g.][]{ma09,kienreich09}.
It is worth noting that many of the EUVI wave studies revealed a close association of the wave and the
erupting CME and its expanding flanks \citep[e.g.][]{veronig08,attrill09,kienreich09,patsourakos09b},
whereas the associated flares were all very weak.
Stereoscopic EUVI studies revealed that the wave signal is typically confined to about 1--2 coronal scale heights above the
chromosphere \citep{patsourakos09,patsourakos09b,kienreich09}.

In this letter, we present the first observations of the full dome of the wave observed in EUV, which is consistent with the 
three-dimensional expansion of a coronal shock front. 
We note that part of a wave dome had been observed in soft X-rays by Yohkoh/SXT for a fast coronal wave that occurred in association with a Moreton wave \citep{narukage04}. We also study the evolution of the wave pulse characteristics (velocity, amplitude, width) in high-cadence EUV imaging over a propagation distance of more than $1\,R_\odot$, and discuss the implications for the physical processes involved.

\section{Data}

The EUVI instrument is part of the Sun Earth Connection Coronal and Heliospheric Investigation
\citep[SECCHI;][]{howard08} instrument suite onboard the STEREO-A(head) and STEREO-B(ehind) spacecraft. 
EUVI observes the chromosphere and low corona in four spectral channels (He~{\sc ii} 304~{\AA}: $T \sim 0.07$~MK, Fe~{\sc ix} 171~{\AA}:
$T \sim 1$~MK, Fe~{\sc xii} 195~{\AA}:  $T \sim 1.5$~MK, Fe~{\sc xv} 284~{\AA}: $T \sim 2.25$~MK) out to 1.7\,$R_\odot$ with a pixel
limited spatial resolution of 1.6$''$/pixel \citep{wuelser04}.
On 2010 January 17, STEREO-B was 69.2$^\circ$ behind Earth on its orbit around the Sun, observing a large-scale coronal wave 
in its Eastern hemisphere. 
The EUVI-B imaging cadence was 2.5 min in the 171~{\AA}, 5 min in the 195~{\AA}, 2.5--5 min in the 284~{\AA}, and
5 min in the 304~{\AA} passband. The EUVI data were reduced using the secchi$\underline{~}$prep routine available within Solarsoft, and corrected for solar differential rotation before we derived base and running difference and ratio images, respectively. We also show white light observations from the COR1-B coronagraph,
which has a FoV of 1.5--4~R$_\odot$, and observed with a cadence of 5~min during the event under study.

\section{Results}

\begin{figure}[p]
\resizebox{16.5cm}{!}{\includegraphics{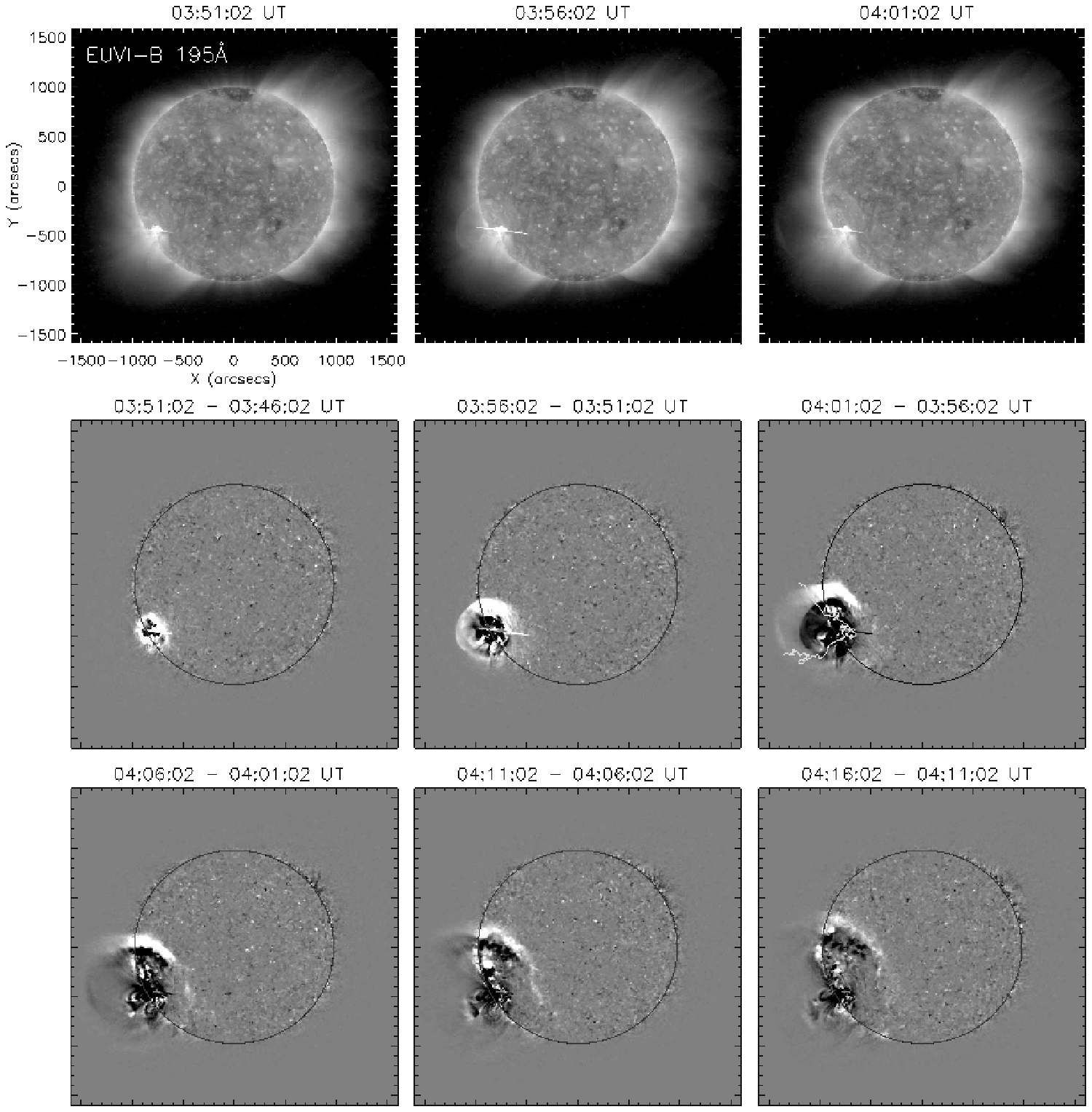}}
\caption{Sequence of STEREO-B EUVI 195~{\AA} images showing the early wave evolution on the disk and above the limb
(top panels: direct images, middle and bottom images: 5-min running difference images). [See also accompanying movie no.~1.]
Note the dome shape in the images at 03:56 and 04:01 UT. In the sixth panel, we overlay the outer contours of the coronal dimming. 
} \label{fig1}
\end{figure}

\begin{figure}[p]
\resizebox{16cm}{!}{\includegraphics{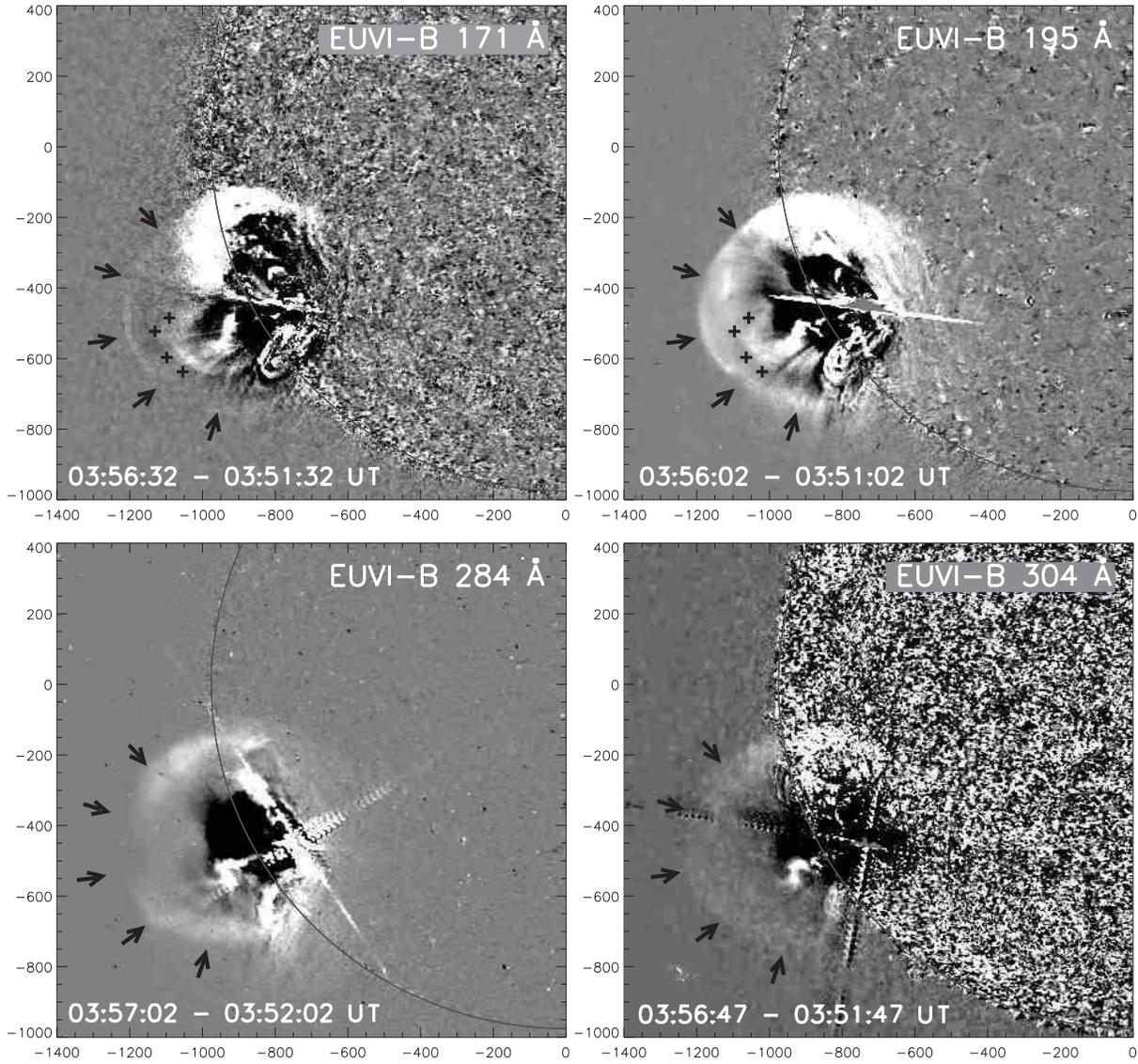}}
\caption{5-min running difference images of the wave dome in all four EUVI-B spectral channels. Arrows outline the wave dome; crosses indicate the erupting CME loops inside the dome.
} \label{fig2}
\end{figure}

Figure~\ref{fig1} shows the evolution of the wave in EUVI 195~{\AA} direct and running difference images, where we subtracted from each frame the
frame taken 5~min before (see also the accompanying movie~no.~1). The wave is best observed in the EUVI 195~{\AA} filter, which has a
broad temperature response peaking at $\sim$1.5~MK, but can be observed in all four wavelengths.
In the 195~{\AA} images at 03:56 and 04:01~UT the full dome of the wave is clearly observed, even
in the direct images. The image sequence also reveals that the on-disk signatures of the wave perfectly connect  to the wave dome
observed above the limb. The sharp and very regular edges of the dome
further suggest that we really observe the shock front of the wave. 
In EUVI 171~{\AA} images, erupting CME loops are observed behind/inside the dome (cf.\ Fig.~\ref{fig2}).

In Fig.~\ref{fig2}, we plot difference images taken almost simultaneously (between 03:56 and 03:57~UT)
in all four EUVI channels. The wave dome can be identified in all four wavelengths, which
indicates that structures at different temperatures are disturbed by the wave, covering at least the temperature range 1.0--2.3 MK.
The fact that we can observe the wave dome also in the EUVI 304~{\AA} filter actually indicates that
Si~{\sc xi} is significantly contributing to the 304~{\AA} emission in addition to the $10^4$~K emission due to He~{\sc ii} lines \citep[see also][]{long08,patsourakos09}.

\begin{figure}[p]
\resizebox{16.5cm}{!}{\includegraphics{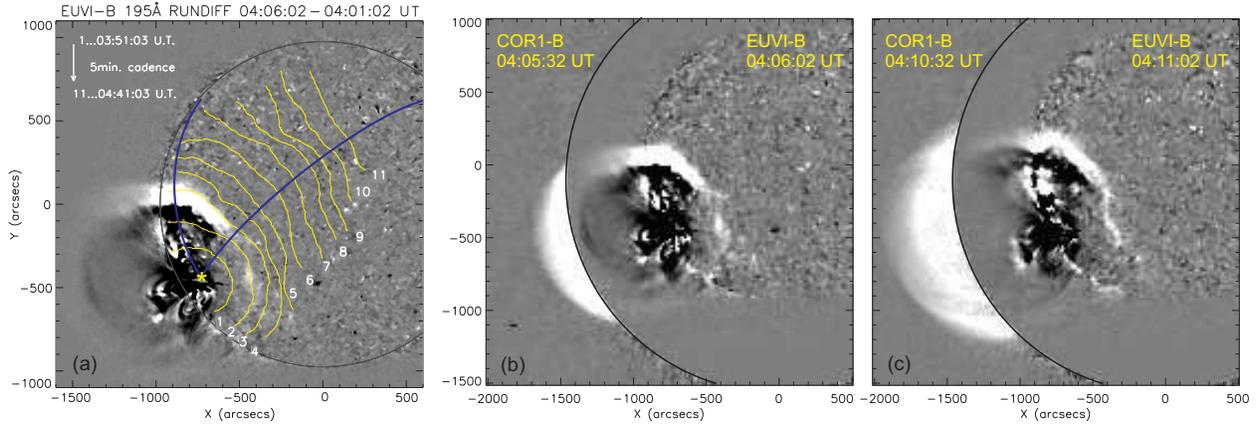}}
\caption{a) Median-filtered EUVI-B 195~{\AA} 5-min running difference image at 04:06~UT. The yellow curves indicate the outer 
edges of the wavefronts visually identified in the EUVI-B 195~{\AA} filtergrams. The blue curves outline the wave propagation 
sector in which the intensity profiles plotted in Fig.~\ref{fig5} were calculated, starting from the determined wave center (indicated by a yellow star).
b), c) Composits of EUVI-B and COR1-B running difference images (note the larger FoV compared to panel a). 
The black circle indicates the inner border of the COR1 FoV at 1.5~$R_\odot$.
} \label{fig3}
\end{figure}

We also note that HIRAS (Hiraiso Radio Spectrograph) reported an associated high-frequency type~II burst drifting from
$\sim$310~MHz to $\sim$80~MHz during $\sim$03:51--03:58~UT. The wave center was derived at a meridional distance of 57$^\circ$ for STEREO-B (cf. Fig.~\ref{fig3}), which
implies that it was located $36^\circ$ behind the Eastern solar limb for an Earth-based vantage point, 
corresponding to an occultation height of about 0.23~$R_\odot \sim 160$~Mm.
Since the radio source was behind the solar limb when looking from the Earth, the observed emission has to be 
at the harmonic of the plasma frequency. 
Applying two- to ten-fold Saito density models this corresponds to the height range
0.11--0.35\,R$_\odot$, 
suggesting that the shock occurred relatively low in the solar corona. These shock formation heights are
consistent with the heights of the wave dome observed in EUV (cf.\ Fig.~\ref{fig1}).

In the sixth panel of Fig.~\ref{fig1}, the outer contours of the coronal dimming region as identified in 195~{\AA} base ratio images (05:01\,UT/03:36\,UT) are overlaid.
At the time, at which we extract the contours, the dimming was maximally developed and darkest; the contour lines plotted in Fig.~\ref{fig1} are at 80\% of the pre-event intensity. It is evident that the North-South extent of the dimming region, which outlines the lateral extent of the CME structure low in the corona (CME flanks),
is significantly smaller than that of the expanding wave dome.
If the dome corresponded to the CME body, its extent should not exceed that of the coronal dimming.
In Fig.~\ref{fig3} (panels b and c) we show composits of EUVI~195~{\AA} and COR1 images. The wave dome observed in EUVI images smoothly extends to the white-light structure
observed in COR1. This suggests that the outer edge observed in the white-light coronagraphic images correspond rather to the coronal shock ahead of the CME
than to the CME leading edge itself \citep[see also][]{vourlidas03}.

\begin{figure}[p]
\resizebox{14cm}{!}{\includegraphics{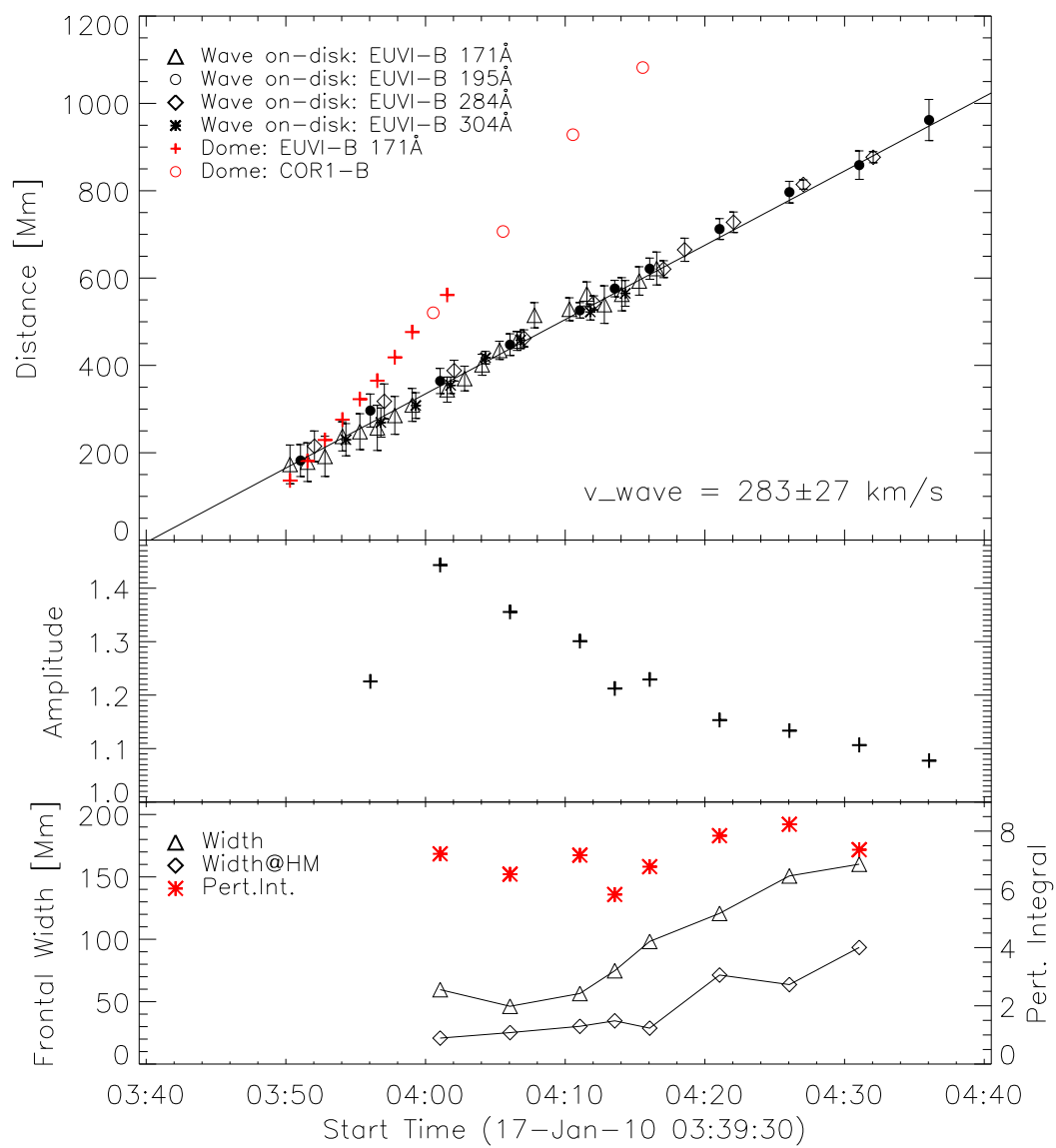}}
\caption{Top: Kinematics of the wavefronts observed on the solar disk in all four EUVI-B channels together with
the upward motion of the wave dome measured in EUVI-B and COR1-B. Middle: Evolution of the perturbation amplitude
determined from the EUVI-B 195~{\AA} intensity profiles shown in Fig.~\ref{fig5}.
Bottom: Evolution of the width (at half maximum: diamonds, full: triangles)
and the integral (stars) of the frontal part of the perturbation profiles.
} \label{fig4}
\end{figure}

Figure~\ref{fig3}a shows an EUVI-B 195~\AA\ difference image together with all wavefronts determined in the 195~\AA\ passband
and the center obtained from circular fits to the earliest wavefronts in the 3D spherical plane \citep[cf.][]{veronig06}.
For each wavefront visually identified in the 195, 171, 284 and 304~{\AA} difference images,
we determined the mean distance from the derived center along the spherical solar surface.
The top panel in Fig.~\ref{fig4} shows the resulting wave kinematics.
The velocity of the wave obtained from the linear fit to the kinematical curve is $v \sim 283\pm 27$~km~s$^{-1}$,
and remains constant over the propagation distance up to 950~Mm. 
In the same panel we also plot the kinematics of the wave dome followed along its main propagation direction as observed in EUVI and COR1.
The distance of the wave dome is measured against the plane-of-sky, and for the starting point we use the center derived for the on-disk EUVI wave measurements.
The plot shows that at the beginning the distance of the on-disk wave and the height of the dome are roughly in agreement,
but the upward movement of the wave dome is much faster ($v \sim 650$~km~s$^{-1}$) than the lateral expansion of the wave observed on the solar disk
($v \sim 280$~km~s$^{-1}$).

In Fig.~\ref{fig5} we plot the evolution of the intensity amplitudes of the wave, so-called ``perturbation profiles'', determined from 195~\AA\ ratio images, where we divided each frame by the frame recorded 10 min before. We calculated the intensity profiles over a 60$^\circ$ sector on the solar sphere (indicated in Fig.~\ref{fig3}a), where the signal of the wave is strongest, by averaging the intensities of all pixels in ``rings'' of increasing radius around the wave origin shown in Fig.~\ref{fig3} \citep[cf.][]{podladchikova05,muhr10}.
Base ratio images would in principle be better suited to study the perturbation profiles. However, these are affected by changes in the
quiet Sun as well as by brightenings induced by the wave front passage that are only slowly fading \citep[``stationary brightenings'';][]{attrill07}.
As a result of several tests, we use 10-min running ratio images since they well represent the
propagating wave profile and ensure that the peak of the wave amplitude is not cut. 

The propagation of the wave is well reflected in the perturbation profile evolution plotted in Fig.~\ref{fig5} (see also the accompanying movie~no.~2).
We also observe steepening and amplitude increase in the profiles 03:56 to 04:01~UT, where the highest amplitude of 1.45 is reached, corresponding to an enhancement of 45\% above the pre-event level. Assuming that the intensity enhancement is primarily due to plasma compression rather than due to temperature changes, which is somewhat justified by the observations of the wave over a broad temperature range in the four EUVI channels, this intensity amplitude of $I/I_0=1.45$ in an optically thin emission line corresponds to a density ratio $n/n_0 \propto (I/I_0)^{1/2} \sim 1.2$.
The perturbation amplitudes in the other EUVI channels are smaller than in 195~{\AA} but clearly recognized; see the
peak amplitudes in 171~{\AA} and 284~{\AA} plotted in the second panel of Fig.~\ref{fig5}.
We also note that at the time of maximum amplitude, the outer edge of the wave front is steepest.
In the subsequent evolution, the amplitude and the steepness of the wave front decrease until it
can no longer be followed in the profiles at about 04:36~UT.

In the middle panel of Fig.~\ref{fig4} we plot the evolution of the intensity amplitude, which decreases continuously after the peak at 04:01~UT.
We fitted the profile amplitude~$A = (I-I_0)/I_0$ as a function of the propagation distance~$d$ with a power-law
$A = a \cdot d^{b}$, giving $b=-2.5 \pm 0.3$. 
We also measured the width of the wave and followed its evolution. Due to the problematics of the stationary brightenings behind the
wave front (which masks the trailing part), we extract only the width of the frontal part, i.e.\ from the profile peak to the outer front.
We derived the full width of the frontal part (defined from the peak down to 2\% enhancement above the pre-event level), as well as the frontal width at half maximum.
These are plotted in the bottom panel of Fig.~\ref{fig4} together with the 
integral of the perturbation profile (as shown in Fig.~\ref{fig5}) over the frontal part of the wave.
The width of the wave pulse increases during its evolution by a factor of 3--4, whereas the integral 
remains basically constant (changes are $\lesssim$25\%).

\section{Discussion and Conclusions}

There are four main arguments supporting the conclusion that the observed EUV structure is the wave dome and not the CME body: 
a) The dome appears spherically three-dimensional with a sharp outer edge; 
inside the dome erupting CME loops are observed. 
b) The wave signatures observed in the low corona above the limb perfectly connect to the wave signatures observed against the solar disk. 
c) The lateral extent of the expanding dome is much larger than that of the coronal dimming.
d) The event was associated with a high-frequency type II burst
indicating shock formation low in the corona, which is consistent with the observed EUV dome height.

The velocity derived for the upward expansion of the wave dome  ($v \sim 650$~km~s$^{-1}$) is larger than that of the lateral expansion of
the wave observed on-disk ($v \sim 280$~km~s$^{-1}$). 
There are at least two alternative explanations for this. In the freely propagating phase, the upward-lateral velocity difference  
can be due to differences of the fast magnetosonic velocities of the active region (AR) and the ambient quiet corona. 
The tip of the wave dome lies above the AR, and the magnetosonic speed is much higher above ARs than at low heights in its surroundings 
\cite[e.g.][]{warmuth05b}. An alternative explanation is that the upward dome expansion is driven all the time, and thus depends on the 
CME speed, whereas in the lateral direction the wave is freely propagating as soon as the lateral expansion of the CME flanks has stopped, 
and its velocity is determined by the characteristic speed of the medium.
The second interpretation is supported by the evolution of the dimming region in the low corona, which is fast expanding up to about 04:01--04:06~UT. Thereafter,
the coronal dimming still gets darker until about 05:00~UT, indicating ongoing mass depletion, but with only little (or no) further expansion after 04:06~UT. 

The high-cadence EUV observations together with the distinct wave signal observed in the low corona against the solar disk allowed us to study
in detail the evolution of the perturbation characteristics during the coronal wave propagation. 
We find that the amplitude of the perturbation first increases and the perturbation profile
steepens within the first 5~min of the event (i.e.\ until 04:01~UT), thereafter the amplitude~$A$
decreases, following a power-law of the form $A \propto r^{-2.5 \pm 0.3}$. The width of the
perturbation profile broadens during its evolution by a factor of 3--4. This broadening is observed
for both the full width as well as for the width at half maximum. 
We stress that we only measured the frontal width of the wave pulse due to the uncertainties in the trailing part (mostly due to stationary brightenings induced by the wave passage). Since the perturbation profile is not necessarily symmetric, the evolution of the total width of the wave pulse may be different,
but we do not expect that it would alter the main outcome of broadening, which is a quite pronounced effect.
The integral below the frontal part of the perturbation profile basically remains constant. 
These characteristics of the wave pulse evolution are all consistent with the spherical expansion and decay of a large-amplitude fast-mode MHD wave.

\begin{figure}[p]
\resizebox{6.8cm}{!}{\includegraphics{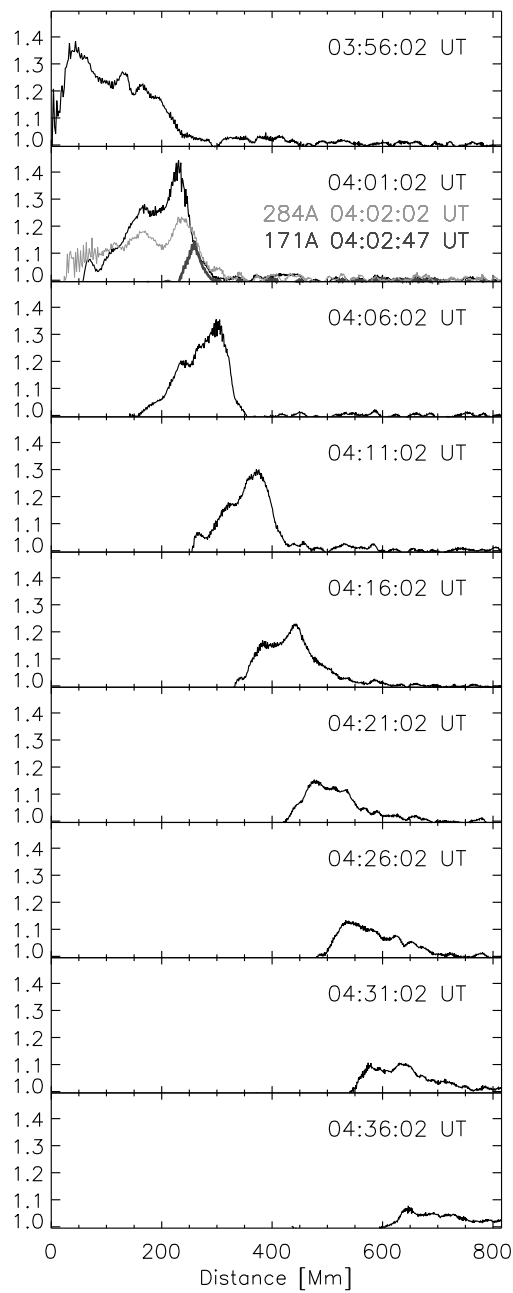}}
\caption{195~{\AA} intensity enhancements (``perturbation profiles'') of the propagating wave derived over a 60$^\circ$ sector where
the wave is strongest (cf.\ Fig.~\ref{fig3}a).
Note that all values behind the propagating wave front which are smaller than one (as a result of the running ratio procedure) 
are set to one. In the second panel, we plot also the peak intensity profiles derived from the
EUVI-B 171 and 284~{\AA} images. [In the accompanying movie no.~2, we show the wave evolution in the EUVI-B 195~{\AA} 10-min running ratio images
together with the corresponding perturbation profiles.] 
} \label{fig5}
\end{figure}

The velocity of the lateral wave expansion ($v \sim 280$~km/s) lies in the range of the fast magnetosonic speed in the quiet solar corona during solar minimum condition \citep[see discussion in][]{kienreich09}. From the observed EUV wave intensities at the highest amplitude reached at $\sim$04:01~UT, we obtained a rough estimate of the peak density jump at the wave front, $n/n_0 \sim 1.2$. This corresponds to a perpendicular fast magnetosonic wave number of $M_{fms} \sim 1.15$ \citep{priest82}.
Thereafter, as the wave decays to $n/n_0=1.04$ (Fig.~\ref{fig5}; last profile at 04:36~UT), it approaches a linear regime since $M_{fms} \sim 1.03$.
These numbers indicate the evolution of a weak shock, and may explain why we do not observe a significant deceleration of the free-propagation of the lateral wave expansion
in association with the amplitude decay.

\acknowledgments AMV, NM, and IWK acknowledge the Austrian Science Fund (FWF): P20867-N16. 
The European Community's Seventh Framework Programme
(FP7/2007-2013) under grant agreement no.~218816 (SOTERIA) is acknowledged (BV, MT).
We thank the STEREO/SECCHI teams for their open data policy.

\bibliographystyle{apj}

\end{document}